\documentclass[twocolumn,showpacs,preprintnumbers,amsmath,amssymb,floatfix]{revtex4}
\usepackage{graphicx}

\begin{document}

\title{Time-dependent universal conductance fluctuations and coherence in AuPd and Ag}

\author{A. Trionfi, S. Lee, and D. Natelson}

\affiliation{Department of Physics and Astronomy, Rice University, 6100 Main St., Houston, TX 77005}

\date{\today}


\begin{abstract}
Quantum transport phenomena allow experimental assessment of the phase
coherence information in metals.  We report quantitative comparisons
of coherence lengths inferred from weak localization magnetoresistance
measurements and time-dependent universal conductance fluctuation
data.  We describe these two measurements and their analysis.  Strong
agreement is observed in both quasi-2D and quasi-1D AuPd samples, a
metal known to have high spin-orbit scattering.  However, quantitative
{\it disagreement} is seen in quasi-1D Ag wires below 10 K, a material
with intermediate spin-orbit scattering.  We consider explanations of
this discrepancy, with particular emphasis on the theoretical
expressions used to analyze the field dependence of the conductance
fluctuations.  We also discuss the mechanism of the suppression of
conductance fluctuations at high drive levels, and dephasing
mechanisms at work in these systems.

\end{abstract}

\maketitle

\section*{I. Introduction} 

Quantum phase coherence in normal metals gives rise to numerous
corrections to the classically predicted conductivity.  These
corrections are commonly referred to as quantum transport phenomena
(QTP).  The study of QTP has both fundamental physical
importance\cite{Imrybook97} and possible implications in novel device
architectures\cite{LiangetAl01Nature,CapassoetAl96JMP}.  The typical
experimental application of QTP is to infer quantum coherence time and
length scales.  Weak localization magnetoresistance
(WL)\cite{Bergmann84PR}, magnetic field-dependent universal
conductance fluctuations\cite{SkocpoletAl86PRL,WashburnetAl92RPP}
(MFUCF), time-dependent universal conductance
fluctuations\cite{BirgeetAl89PRL,Giordanobook}(TDUCF), and
Aharonov-Bohm oscillations\cite{WebbetAl85PRL} have all been used to
examine coherence in normal metals.

QTP in diffusive conductors arise due to interference between possible
electronic trajectories from one location to another.  Interference,
however, is only relevant when the phase of the conduction electrons'
partial waves is well-defined.  The coherence length is defined as the
distance scale over which the phase of a conduction electron's wave
function remains correlated to its initial phase.  This length can be
related to a coherence {\it time}, $\tau_{\phi}$, by $L_{\phi} =
\sqrt{D \tau_{\phi}}$ where $D$ is the diffusion constant of the
electron in the disordered solid.  Decoherence or dephasing can occur
when an electron experiences an interaction with another dynamical
degree of freedom.  The three most common dephasing processes are
electron-electron scattering, electron-phonon scattering, and
spin-flip interactions with magnetic impurities.  The rates of these
processes have distinct temperature dependences, allowing QTP to be
used to distinguish between these mechanisms in various metals.


Interesting questions have arisen from experimental characterization
of electron coherence.  One question is whether precisely the same
coherence length is inferred from different QTP.  This is a subtle
issue because the precise time scales and processes relevant to a
particular observable can be complicated, and the evolution of
electronic phase correlations is typically not a simple single-time
exponential decay.  A previous test of this coherence length
consistency led to equivocal results in quasi-2D
silver\cite{HoadleyetAl99PRB}.  Another question is the cause of an
observed low temperature saturation of the coherence length in many
materials\cite{MohantyetAl97PRL}.  An explanation with significant
experimental support is scattering from dilute concentrations of low
Kondo temperature magnetic impurities\cite{PierreetAl03PRB}, while
others suggest intrinsic mechanisms\cite{GolubevetAl98PRL}.  These two
questions are increasingly related: Recent
publications\cite{PierreetAl02PRL,MohantyetAl03PRL} have compared
experimental results from different QTP when debating the cause of
coherence saturation; it must be established, however that these
analyses are truly comparing equivalent parameters.

In this paper we briefly review the physics underlying WL, MFUCF, and
TDUCF, and report measurements of these effects in two different
materials, Au$_{0.6}$Pd$_{0.4}$ and Ag, over a broad temperature and
field range.  While we find excellent quantitative agreement between
$L_{\phi}^{\rm WL}(T)$ and $L_{\phi}^{\rm TDUCF}(T)$ in all AuPd
samples, we observe a {\it divergence} between these two inferred
coherence lengths in quasi-1d Ag samples, as seen previously in
quasi-2D Ag films\cite{HoadleyetAl99PRB}.  We discuss candidate
explanations, and suggest that a likely concern is the applicability
of the theoretical expressions used to analyze the TDUCF field
dependence.  We also show that the suppression of TDUCF amplitude at
high drive currents is consistent with bias-induced energy averaging.
Finally we discuss the implications of these data on dephasing
mechanisms at work in these systems.

Weak localization arises from the properties of electronic
trajectories under time-reversal symmetry.  Many electronic paths in a
diffusive conductor contain loops.  Without a magnetic field, an
electron circumnavigating such a loop accumulates the same phase as
one doing so under time-reversed conditions.  This phase agreement
causes constructive interference that enhances back-scattering,
leading to a conductivity lower than is classically predicted.  With
strong spin-orbit interactions, the sign of this interference is
reversed and leads to enhanced conduction.  In the presence of a
magnetic field normal to the loop, the vector potential adds opposite
phase shifts to each looped path and corresponding time-reversed
conjugate.  This eliminates the constructive (destructive)
interference when $\sim$ one quantum of flux is threaded through a
typical loop.  The result is a magnetoresistance with a field scale
related to the area of a typical coherent loop, allowing inference of
$L_{\phi}^{\rm WL}$.

Time-dependent UCF are due to the enhanced sensitivity of the
conductance to the motion of individual scatters.  Unlike weak
localization, all interfering paths contribute to this phenomenon.
When a scattering site moves, it changes the interference
pattern of all intersecting electronic paths within a coherent volume
of the scattering site, leading to a conductance change.  A single
moving scatter can change the conductance within a coherent volume at
zero temperature by roughly $e^{2}/h$.  If the relaxation times of the
scatters are appropriately distributed, the TDUCF exhibit a $1/f$
power spectrum\cite{FengetAl86PRL}, which is the case in many normal
metals.  Much like weak localization, the noise power of the $1/f$
noise is sensitive to a perpendicular magnetic field.  The
time-reversed loop contribution (the cooperon) will be suppressed as
the field is increased while the sensitivity due to all remaining
paths, known as the diffuson, remain unchanged\cite{Stone89PRB}.  This
leads to a factor-of-two decrease in the noise power as the field is
ramped up, and allows extraction of $L_{\phi}^{\rm TDUCF}$.

Magnetic field-dependent UCF are closely related to the time-dependent
form.  The explanation of this phenomenon comes from the ergodic
hypothesis\cite{LeeetAl85PRL,AltshuleretAl86JETPL}, which implies that
other effects that randomize the interference of electronic paths are
equivalent to scattering site motion.  Since a perpendicular magnetic
field introduces an Aharonov-Bohm phase shift particular to each
electronic path, varying such a field leads to conductance
fluctuations of the universal size of $\sim e^{2}/h$.  The result is a
completely reproducible magnetoresistive pattern that is
sample-specific, commonly referred to as the magnetofingerprint.

Although the expected size of the conductance fluctuations is
universal, the measured effect may be much smaller.  Samples much
longer and wider than $L_{\phi}$ may be treated as uncorrelated
fluctuators in series and parallel.  The measured noise power is
therefore reduced by a factor $N$, the number of coherent volumes
between the ends of the measured sample.  Further averaging occurs
when the energy range of accessible single-particle states exceeds the
correlation energy\cite{Stone85PRL}, $E_{\rm c}=\hbar D/L_{\phi}^{2}$.
In this case the relevant states can be subdivided by energy into
coherent sub-bands, each nominally uncorrelated with the others,
leading to further ensemble averaging.  One way to increase available
energy levels is via thermal energy.  The thermal length, $L_{T}\equiv
(\hbar D/k_{\rm B}T)^{1/2}$, is the distance two initially in-phase
electrons separated in energy by $k_{\rm B}T$ may move before their
phases differ by $\sim 1$.  The condition $L_{T}< L_{\phi}$ is
equivalent to $k_{\rm B}T > E_{\rm c}$, leading to ensemble averaging
by $L_{T}/L_{\phi}$ to some power.  Similarly, when $eV_{\rm c}>
E_{\rm c}$, where $V_{\rm c}$ is the voltage dropped across a
coherence length, ensemble averaging will also
occur\cite{WashburnetAl92RPP}.


The observed magnitude of the TDUCF may also be smaller than the
universal limit if the conductance fluctuations are not ``saturated''.
A sample is said to be in the saturated regime if the conductance
variance within a coherent volume has reached the limiting $\sim
e^{2}/h$ amplitude.  How close a sample is to this condition depends
on the microscopic nature of the fluctuators.  Since MFUCF should
always exhibit conductance fluctuations on order $e^{2}/h$ within a
coherent volume, if a sample is in the saturated regime, the TDUCF
noise power integrated over the bandwidth of the fluctuators should
equal the MFUCF magnitude.  That is, $\int_{0}^{\infty}S_{G}(f)df$
should $=(\delta G^{\rm MFUCF})^{2}$, where $S_{G}$ is the conductance
noise power, $f$ is the frequency, and $G$ is the conductance.  If the
fluctuators are typical tunneling two-level systems (TLS) of the type
ubiquitous in disordered solids\cite{Esquinazibook}, their relaxation
rates are estimated to span $\sim$ 20 decades\cite{BirgeetAl89PRL}.
TDUCF measurements in the literature are all thought to be
nonsaturated.  This issue is important, as it determines the
functional form appropriate for analysis of the field dependence of
the TDUCF.

For WL and TDUCF, the number of the sample dimensions longer than
$L_{\phi}$ determines the effective dimensionality of the system with
regard to coherence effects.  Thus a quasi-2d sample is achieved when
$t<<L_{\phi}<<w$ and a quasi-1d sample when $w,t << L_{\phi}$.  Here
$t$ and $w$ are sample thickness and width, respectively.

Previous comparisons between $L_{\phi}^{\rm WL}(T)$ and $L_{\phi}^{\rm
TDUCF}(T)$ have shown a disagreement between these parameters at low
temperatures in quasi-2d Ag samples\cite{HoadleyetAl99PRB}.  This was
interpreted as evidence supporting a theoretical
treatment\cite{Blanter96PRB} that argued that the Nyquist or
electron-electron dephasing rate would limit the coherence in WL while
the out-scattering rate would limit the coherence in universal
conductance fluctuations.  The out-scattering rate is the rate at
which an electron will change its momentum state in the Boltzmann
formalism.  It was shown the two rates have different temperature
dependencies at low temperature so a divergence between the coherence
lengths inferred from WL and TDUCF was expected.  Recent corrections
to the theory\cite{AleineretAl02PRB} show that, as long as
electron-electron scattering is the only small-energy-transfer
inelastic process, $L_{\phi}^{\rm WL}$ is expected to equal
$L_{\phi}^{\rm TDUCF}$.  This leaves the experimental results in Ag
without an explanation.  We discuss this further below.

\section*{II.  Procedure}

Samples were patterned on undoped GaAs substrates using standard
electron beam lithography, as discussed in
Ref.~\cite{TrionfietAl04PRB}.  Distances between consecutive leads
(voltage or current) were 10~$\mu$m in AuPd samples and 20~$\mu$m in
Ag samples.  Ag (0.99999 purity) samples were made using a single
lithography/deposition step with the wire and leads all silver, and no
adhesion layer.  Au$_{0.6}$Pd$_{0.4}$ samples were made using two
lithography steps, the first for the AuPd wire and the second for the
Ti/Au leads (1.5~nm Ti, 25~nm Au).  The AuPd source material is
expected to be free of ferromagnetic impurities to the 10$^{-5}$
level.  To minimize the contact resistance between the AuPd wires and
the Ti/Au leads, samples were exposed to oxygen plasma for 30 seconds
prior to the Ti/Au deposition to remove any resist residue.  Typical
contact resistances in the AuPd samples were less than 30~$\Omega$.
All depositions were performed using an electron beam evaporator at $5
\times 10^{-7}$~mB.  To test the effects of magnetic impurities on the
consistency of $L_{\phi}^{\rm WL}$ and $L_{\phi}^{\rm TDUCF}$, one
AuPd sample was deliberately contaminated with trace impurities by
evaporating 2.5 nm of Ni$_{0.8}$Fe$_{0.2}$ with the sample shutter
{\it closed}, prior to the AuPd evaporation.

\begin{table}
\caption{Samples used in magnetotransport and noise measurements.
Free electron density of states for Au and Ag used to calculate $D$
for AuPd and Ag samples: $1 \times 10^{47}$~m$^{-3}$J$^{-1}$, from
Ref.~\protect{\cite{Ashcroft75}}.  Diffusion constants calculated via the Einstien relations.  Sample D was deliberately
contaminated with additional ferromagnetic impurities as described in
the text.  The effective dimensionality $d$ for coherence effects
is determined by the relative size of $L_{\phi}$ and the sample
dimensions.}
\begin{tabular}{c c c c c c c}
\hline \hline
Sample & metal & $d$ & $w$ & $t$ & $R/L$ [$\Omega$/$\mu$m] (1d) & $D$   \\
& & & [nm] & [nm] & $R/\Box$ [$\Omega$] (2d) & [m$^{2}$/s]\\
\hline
A & AuPd & 1 & 43 & 9 & 722 & 1.34 $\times 10^{-3}$ \\
B & AuPd & 1 & 35 & 9 & 857 & 1.34 $\times 10^{-3}$ \\
C & AuPd & 2 & 500 & 6.5 & 84.5 & 7.9 $\times 10^{-4}$ \\
D & AuPd & 2 & 500 & 8.5 & 47.5 & 9.6 $\times 10^{-4}$ \\
E & Ag & 1 & 115 & 12 & 49 & 5.65 $\times 10^{-3}$ \\
F & Ag & 1 & 140 & 12 & 35 & 6.70 $\times 10^{-3}$ \\
G & Ag & 1 & 130 & 12 & 45 & 5.63 $\times 10^{-3}$ \\
H & Ag & 1 & 70 & 20 & 42 & 6.50 $\times 10^{-3}$ \\
I & Ag & 1 & 125 & 12 & 43 & 5.91 $\times 10^{-3}$ \\
J & Ag & 1 & 100 & 12 & 88.5 & 3.61 $\times 10^{-3}$ \\
\hline
\hline
\end{tabular}
\label{tab1}
\end{table}

Measurements between 2 and 20~K were performed in a $^{4}$He cryostat
while lower temperatures for two samples (G,H) were achieved in a
dilution refrigerator.  All samples were initially characterized with
standard ac four terminal resistance measurements and tested as a
function of temperature at various drive currents to check for Joule
heating.  Upturns in the resistance at low temperatures were
consistent with electron-electron interaction corrections.  All
subsequent measurements were performed at or below the limiting
current set by the Joule heating tests. 

WL magnetoresistance measurements were made using standard four
terminal techniques.  The applied field was swept between $\pm$1.25~T
for AuPd samples while the field range for the Ag samples was $\pm$
0.9~T.  The TDUCF measurement employed an ac five terminal bridge
technique developed by Scofield\cite{Scofield87RSI}.  The ac bridge
technique renders the noise measurement insensitive to noise in the
voltage source.  The bridge was measured with a low noise differential
pre-amplifier (1.5~nV/$\sqrt{\rm Hz}$, NF Electronics LI-75A).  Signal
frequencies ranged from 600~Hz for AuPd samples to 1~kHz for Ag
samples, chosen to optimize the noise contours of the preamplifier.
Both the in-phase and out-of-phase demodulated signals were fed into a
two-channel dynamic signal analyzer (SRS SR785) to transform the data
into the frequency domain.  Strong $1/f$ dependent spectra were
consistently observed from the in-phase channel while the out-of-phase
channel provided a measure of the white background noise of the
measurement circuit.  By subtracting away this background, the
sample-induced noise could be isolated.  A typical frequency span
ranged from 78~mHz to 1.5~Hz for the AuPd samples and to 3~Hz in Ag
samples.  The low temperature noise power in all samples show the
expected amplitude increase with decreasing temperature as well as the
factor of two drop in the presence of a large perpendicular magnetic
field (except when local interference noise becomes non-negligible).
This observation is consistent with the expected TDUCF behavior.



\begin{figure}[h!]
\begin{center}
\includegraphics[clip, width=8cm]{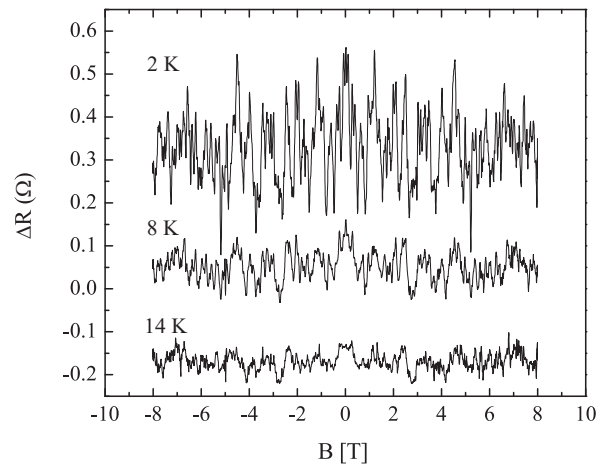}
\end{center}
\vspace{-3mm}
\caption{``Magnetofingerprint'' measurements on sample I made using the five terminal bridge technique.  The curves from top to bottom are at 2, 8, and 14 K.  The curves are offset for clarity.}
\label{revfig1}
\end{figure}

The bridge technique was also employed to make the MFUCF measurements.
By using this method instead of the standard four terminal resistance
measurement, the WL magnetoresistance is nulled away since both sides
of the bridge will have identical resistance changes due to this
effect.  The nulling of the WL magnetoresistance allows the
magnetofingerprint to be observed down to zero field.  An example
of the magnetofingerprint is given in Figure~\ref{revfig1}.

\section*{III.  Analytical Approach}

\subsection*{A.  Weak localization}
Values of $L_{\phi}^{\rm WL}$ were inferred from
the AuPd WL magnetoresistance using the following equations for
one and two dimensions
respectively\cite{PierreetAl03PRB,LinetAl87PRB}:
\begin{equation}
\frac{\Delta R}{R}\big{\vert}_{\rm 1d} = - \frac{e^{2}}{2 \pi \hbar}\frac{R}{L}\left[\frac{1}{L_{\phi}^{2}}+\frac{1}{12}\left(\frac{w}{L_{B}^{2}}\right)^{2}\right]^{-1/2}
\label{eq:1dwl}
\end{equation}
\begin{equation}
\frac{\Delta R}{R}\big{\vert}_{\rm 2d} = \frac{e^{2}}{4 \pi^{2} \hbar}R_{\Box}\left[\psi\left(\frac{1}{2}+\frac{1}{2}\frac{L_{B}^{2}}{L_{\phi}^{2}}
\right)-\ln\left(\frac{1}{2}\frac{L_{B}^{2}}{L_{\phi}^{2}}\right)\right]
\label{eq:2dwl}
\end{equation}
$\psi$ is the digamma
function and $L_{B}\equiv \sqrt{\hbar/2eB}$.  $R_{\Box}$ is the sheet
resistance of the sample, $R$ is $R(B=\infty)$, $w$ is the sample
dimension transverse to both the applied field and current flow, and
$L$ is the length of the sample parallel to current flow.  
These equations apply in the limit of {\it strong} spin-orbit
scattering ($\tau_{\rm SO} << \tau_{\phi}$).  AuPd has long been
established as a strong spin-orbit scattering
material\cite{LinetAl87PRB}.

The $L_{\phi}^{\rm WL}$ values inferred from the quasi-1d Ag wires used 
the more general form\cite{PierreetAl03PRB}, 
\begin{eqnarray}
\frac{\Delta R}{R}\big{\vert}_{\rm 1d} & =  - \frac{e^{2}}{2 \pi \hbar}\frac{R}{L} \times \nonumber \\ 
& \left[\frac{3}{\left(\frac{1}{L_{\phi}^{2}}+\frac{4}{3 L_{\rm SO}^{2}}+\frac{1}{12}\left(\frac{w}{L_{B}^{2}}\right)^{2}\right)^{1/2}} - \frac{1}{\left(\frac{1}{L_{\phi}^{2}}+\frac{1}{12}\left(\frac{w}{L_{B}^{2}}\right)^{2}\right)^{1/2}}\right]
\label{eq:1dwlso}
\end{eqnarray}
Here $L_{\rm SO}\equiv \sqrt{D \tau_{\rm SO}}$ is the spin-orbit
length.  $\Delta R$ for Eq.~(\ref{eq:2dwl}) is defined as $\Delta
R=R(B)-R(B=0)$, while it is defined as $\Delta R=R(B)-R(B=\infty)$ in
Eqs.~(\ref{eq:1dwl}) and (\ref{eq:1dwlso}).  The only fitting
parameter for the AuPd curves is $L_{\phi}$, while both $L_{\phi}$ and
$L_{\rm SO}$ are free in the fits for the Ag curves.  The width is
left free at 2~K in all one-dimensional fits to confirm sample size
and is then fixed for all subsequent fits.  Spin-orbit lengths are
also fixed above 10~K to the average value found from lower
temperature fits.

\subsection*{B.  TDUCF noise}
$L_{\phi}^{\rm TDUCF}$ values were inferred from fits of the noise
power as a function of perpendicular field to the appropriate
crossover function, $\nu(B)$, the theoretically expected functional
form.  There are two methods of calculating $\nu(B)$; we have
used both approaches and compared the results.  First,
analytical expressions for the
one-dimensional and two-dimensional crossover functions with large
spin-orbit interaction have been derived recently by Aleiner\cite{Aleinerpc}:
\begin{equation}
\nu_{\rm 1d}(B)=1- \frac{x}{2}\left(\frac{Ai(x)}{Ai'(x)}\right)^{2}
\label{eq:1dnoise}
\end{equation}
where $x \equiv L_{\phi}^{2}/(3(\hbar/B e w)^{2})$,
and
\begin{equation}
\nu_{\rm 2d}(B)=\frac{1}{2}+ \frac{L_{B}^{2}}{4 L_{\phi}^{2}}\psi'\left(\frac{1}{2}+\frac{L_{B}^{2}}{2 L_{\phi}^{2}}\right).
\label{eq:2dnoise}
\end{equation}
These functional forms are strictly valid when
$\hbar/\tau_{\phi}<< k_{\rm B}T$.  Here $Ai(x)$ is the Airy
function, and $\psi'(x)$ is the derivative of the digamma function.

Prior to the derivation of these analytical results, the noise
crossover function was calculated from the theoretical expression for
the magnetic field correlation function $F(\Delta E, \Delta B,
B)\equiv \langle \delta g(E_{\rm F},B)\delta g(E_{\rm F}+\Delta E,
B+\Delta B)\rangle$ of the
MFUCF\cite{LeeetAl87PRB,Stone89PRB,BirgeetAl90PRB}.  Here $g$ is the
conductance in units of $e^{2}/h$.  An approximation of this
correlation function has been derived by Beenakker and van
Houten\cite{BeenakkeretAl88PRB} for quasi-1d samples.  We also analyze
the data with this method and compare with the analytical expressions
above.  The samples are assumed to be in the unsaturated regime, so
that the derivative of $F$ with respect to $\tau_{\phi}^{-1}$ 
must be computed, as per the explanation given by Stone\cite{Stone89PRB}.
The resulting derivative has the form:
\begin{equation}
F'(B)=\frac{L_{\phi B}^{5} \left(1 + \frac{3 L_{\phi B}^{2}(B)}{2 \pi L_{T}^{2}}\right)}{\left(1 + \frac{9 L_{\phi B}^{2}(B)}{2 \pi L_{T}^{2}}\right)}
\label{eq:fder}
\end{equation}
where
\begin{equation}
L_{\phi B}^{2}(B) = \frac{3 L_{\phi}^{2}}{(BeL_{\phi} w/\hbar)^{2}+3}.
\end{equation}
Spin-orbit interactions may be accomodated by changing Eq.~(\ref{eq:fder})
to\cite{ChandrasekharetAl90PRB}:
\begin{equation}
F'(B)=\frac{L_{\phi B}^{5} \left(1 + \frac{3 L_{\phi B}^{2}(B)}{2 \pi L_{T}^{2}}\right)}{4 \left(1 + \frac{9 L_{\phi B}^{2}(B)}{2 \pi L_{T}^{2}}\right)}
+ \frac{3L_{\phi Bt}^{5} \left(1 + \frac{3 L_{\phi B t}^{2}(B)}{2 \pi L_{T}^{2}}\right)}{4 \left(1 + \frac{9 L_{\phi B t}^{2}(B)}{2 \pi L_{T}^{2}}\right)}
\label{eq:fderso}
\end{equation}
where
\begin{equation}
L_{\phi B t}^{2}(B) = \frac{3 L_{\phi}^{2} L_{\rm SO}^{2}}{(BeL_{\phi}L_{\rm SO} w/\hbar)^{2}+3 L_{\rm SO}^{2} + 4 L_{\phi}^{2}}.
\end{equation}
The approximate crossover function is therefore
\begin{equation}
\nu(B) = \frac{1}{2} + \frac{F'(B)}{2 F'(B=0)}.
\end{equation}

The exact crossover function in quasi-2d systems was reported by
Stone\cite{Stone89PRB} and was also used to infer $L_{\phi}^{\rm
TDUCF}$ in the quasi-2d AuPd samples.  The fitting method employed was
that described in Ref.~\cite{McConvilleetAl93PRB}.  A comparison
between the crossover functions computed from the correlation
functions and the analytic expressions of
Eqs.~(\ref{eq:1dnoise},\ref{eq:2dnoise}) finds the following results.
In the quasi-1d case, the $L_{\phi}^{\rm TDUCF}$ values inferred using
the correlation functions systematically exceed those extracted using
the analytical expressions by roughly 10\%.  Similarly, in the
quasi-2d case, the correlation function-based values exceed those from
Eq.~(\ref{eq:2dnoise}) by 3\%.

The actual fitting functions used to analyze the normalized noise
power data included an additional fitting parameter to account for the
local interference
noise\cite{DuttaetAl81RMP,PelzetAl87PRB,Hershfield88PRB} that
increases to non-negligible magnitudes at higher temperatures.  Since
local interference noise has no low order field dependence, the noise
power will not drop by a full factor of two at higher temperatures.
The corrected fitting function has the form $f(B) = (1-z) + z\nu(B)$
where $z$ represents the fractional size of the UCF enhanced noise.
Values of $z$ were indistinguishable from 1 for all data sets except
at 20~K in the quasi-2d AuPd samples and quasi-1d Ag samples.  All
fitting was performed using standard $\chi^{2}$ minimization.

\subsection*{C.  Role of magnetic impurity scattering}

As was discussed extensively in Ref.~\cite{PierreetAl03PRB}, magnetic
impurity scattering can affect weak localization and UCF coherence
corrections {\it differently} or {\it identically} depending on the
temperature scale and impurity concentration.  At temperatures higher
than a crossover temperature, the Korringa time for impurity spins to
relax back to thermal equilibrium with the lattice is short compared
to the spin-flip scattering time.  In this regime ($T> T^{*}\equiv \sim 40~{\rm
mK}\times$ the ppm concentration of magnetic impurities for typical
host noble metals), spin-flip scattering should involve large energy
transfers\cite{Aleinerpc}, and affect WL and TDUCF {\it identically}.
At temperatures below this cutoff, spin scattering is more rapid than
the relaxation of the impurity spins; under these conditions, the
spin-flip scattering time is predicted to affect TDUCF and WL {\it
differently}.

An estimate of the decoherence rate due to magnetic impurities may be
obtained from the Nagaoka-Suhl
expression\cite{vanHaesendoncketAl87PRL}:
\begin{equation}
\frac{1}{\tau_{\rm sf}}=\frac{c_{\rm mag}}{\pi \hbar \nu(E_{\rm F})} \frac{\pi^{2}S(S+1)}{\pi^{2}S(S+1)+\ln^{2}(T/T_{\rm K})},
\label{eq:nagaokasuhl}
\end{equation}
where $\nu(E_{\rm F})$ is the density of states at the Fermi level of
the host metal, $S$ is the spin of the impurity, and $T_{\rm K}$ is
the Kondo temperature of the impurity in the host metal.

\section*{IV.  Results and Discussion}

\begin{figure}[h!]
\begin{center}
\includegraphics[clip, width=8cm]{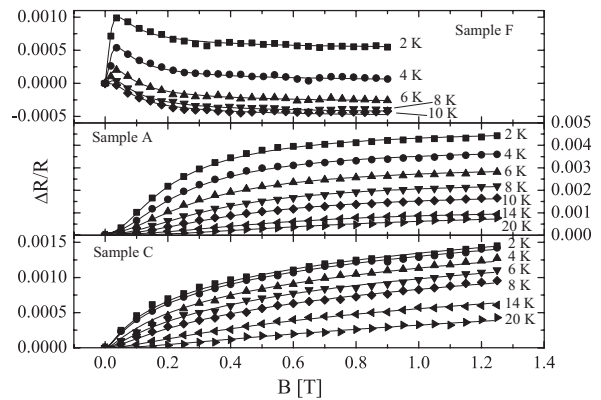}
\end{center}
\vspace{-3mm}
\caption{WL magnetoresistance curves at various temperatures for a 43~nm and 500~nm AuPd wire (samples A and C)  and a 140 nm Ag wire (sample F).  Quasi-1D data shifted
to pass through origin.}
\label{revfig2}
\end{figure}

Magnetoresistance curves for both AuPd and Ag samples are given in
Figure~\ref{revfig2}.  The data are fit very well by
Eqs.~(\ref{eq:1dwl},\ref{eq:2dwl},\ref{eq:1dwlso}).  Sample widths
inferred for the quasi-1d wires via the fitting procedure are
consistent with SEM images and estimates based on resistances of
codeposited films.  Including $L_{\rm SO}$ as a fit parameter in the
AuPd data leads to $L_{\rm SO} \lesssim$~10~nm, with little impact on
$L_{\phi}$.  We find $L_{\rm SO}\approx$~290~nm for Ag, and
$\sim$~9~nm for AuPd.

\begin{figure}[h!]
\begin{center}
\includegraphics[clip, width=8cm]{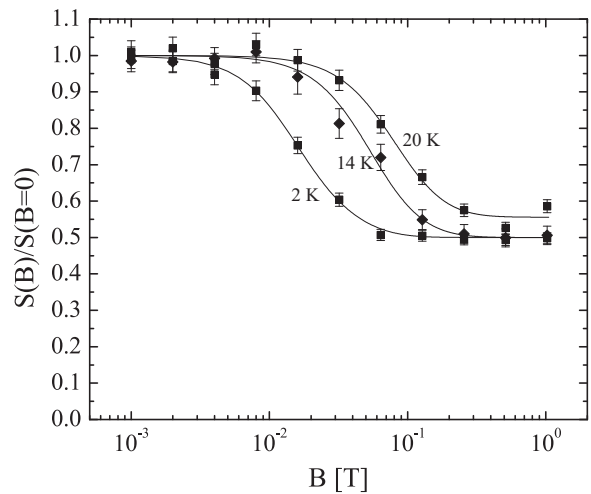}
\end{center}
\vspace{-3mm}
\caption{Noise power data from Ag sample F at 2, 14 and 20~K.  The crossover field becomes larger as the coherence length diminishes.  The 20~K data does not drop by a full factor of 2 due to local interference noise.}
\label{revfig3}
\end{figure}

Similarly, examples of the measured normalized noise power
$(S_{R}(B)/S_{R}(B=0))$ versus field are shown in Figure~\ref{revfig3}
for sample F, a quasi-1d silver wire.  As is clear from the graph,
these data are fit well by the Beenakker/van Houten correlation
function approach (Eq.~(\ref{eq:fderso}) and following).

\subsection*{A.  AuPd comparison of WL and TDUCF}

Figure~\ref{revfig4}a shows the resulting coherence lengths inferred from
both QTP for the two quasi-1d AuPd samples, as originally reported in
Ref.~\cite{TrionfietAl04PRB}.  The inferred $L_{\phi}^{\rm WL}$ and
$L_{\phi}^{\rm TDUCF}$ are in {\it strong quantitative agreement} for
the AuPd samples over the temperature range measured.  

As shown in Ref.~\cite{TrionfietAl04PRB}, this agreement remains
strong even in the presence of magnetic impurity scattering
significant enough to suppress the coherence length by more than a
factor of two.  This strongly supports the theoretical
statement\cite{AleineretAl02PRB} that weak localization and UCF
measurements probe {\it precisely} the same coherence physics, even in
the presence of strong spin-orbit and magnetic impurity scattering
over this temperature range.  Furthermore, the agreement persists even
though $\hbar/\tau_{\phi}$ is never $<< k_{\rm B}T$, suggesting that
Eqs.~(\ref{eq:1dnoise},\ref{eq:2dnoise}) are robust even when that
constraint is somewhat relaxed.

\subsection*{B.  Ag comparison of WL and TDUCF}

\begin{figure}[h!]
\begin{center}
\includegraphics[clip, width=8cm]{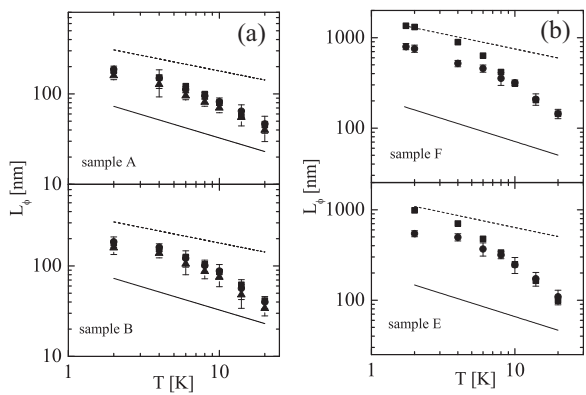}
\end{center}
\vspace{-3mm}
\caption
{(a) Coherence lengths as a function of $T$ for the AuPd samples (lettered accordingly).  Triangles points are inferred from Eq.~(\ref{eq:1dnoise}), circles are inferred from Eq.~(\ref{eq:fder}), and squares are inferred from WL.  Only one UCF fit is shown in quasi-2D since the two results only differ by 3\%.  The dashed lines show the predicted dephasing length due to Nyquist scattering calculated from sample parameters.  The solid line represents the calculated thermal length. (b) Coherence lengths as a function of $T$ for Ag samples E and F (lettered accordingly).  Circle points are inferred from Eq.~(\ref{eq:fderso}), and squares are inferred from WL.  The dashed lines show the predicted dephasing length due to Nyquist scattering calculated from sample parameters.  The solid line represents the calculated thermal length.  Unlike the AuPd case, there is a statistically significant discrepancy between $L_{\phi}^{\rm WL}$ and $L_{\phi}^{\rm TDUCF}$ in these samples. }
\label{revfig4}
\end{figure}

Figure~\ref{revfig4}b shows the equivalent $L_{\phi}(T)$ data for the Ag
samples over the same temperature range.  Note that $L_{\phi}^{\rm
WL}(T)$ shows no saturation, and at low temperatures approaches the
Nyquist predicted value with {\it no} adjustable parameters.  This
strongly suggests that e-e interactions are the only non-negligible
dephasing mechanism in the Ag samples at temperatures near 2 K.  

Comparing Fig.~\ref{revfig4}b with Fig.~\ref{revfig4}a highlights a dramatic
difference between the Ag and AuPd data: There is substantial
disagreement between $L_{\phi}^{\rm WL}$ and $L_{\phi}^{\rm TDUCF}$ in
these quasi-1D Ag samples.  In particular, below 8~K, $L_{\phi}^{\rm
TDUCF}$ is shorter and has a significantly weaker temperature
dependence than $L_{\phi}^{\rm WL}(T)$.  This difference is very
similar to that observed previously in quasi-2D Ag
films\cite{HoadleyetAl99PRB}.

The reason for this disagreement is unknown.  A possibility put
forward by Aleiner and Blanter is that a subtle effect due to triplet
channel electrons is responsible\cite{AleineretAl02PRB}.  This
suggestion reflects the observation in the quasi-2d Ag data that the
disagreement appears at temperatures below $L_{\phi} \approx L_{\rm
SO}$.  Our AuPd data, however, appears {\em inconsistent} with such an
explanation.  The inferred coherence lengths in a strong spin-orbit
scattering material (AuPd) should resemble the low-temperature
limiting behavior of a material with intermediate spin-orbit
scattering (Ag).  The WL/TDUCF agreement seen in the AuPd would imply
that the coherence lengths inferred from the Ag should {\it agree} as
the temperature is reduced.  To the lowest temperatures measured, no
such convergence is observed.  A related prediction\cite{Aleinerpc}
would be for a signature of unusual triplet effects in $R(T)$ when
$k_{\rm B}T < \sim \hbar/\tau_{\rm SO}$.  Since $\tau_{\rm SO} \approx
3.5\times 10^{-11}$~s, this crossover would be predicted at
$\sim$~200~mK.  No change in the $R(T)$ properties is observed down to
50~mK.  TDUCF noise measurements at these lower temperatures have yet
to be performed successfully and are very challenging due to Joule
heating concerns.  On the basis of the AuPd data at
hand, a triplet channel effect seems extremely unlikely to 
explain the differences between Fig.~\ref{revfig4}a and b. 

We suggest another possible resolution to this discrepancy between the
AuPd and Ag data: the applicability of the noise crossover expressions
used in analyzing the data.  Due to microscopic differences in the
(unknown) fluctuators responsible for the TDUCF noise, the AuPd and Ag
samples may be in different regimes.  In particular, if the Ag samples
were transitioning into the {\it saturated} noise limit, then the
fitting functions based on $F'(B)$ above used to infer the coherence
length would be inappropriate.  In the saturated limit, $F(B)$ rather
than its derivative with respect to the inelastic rate is the correct
function from which to derive $\nu(B)$.

Comparing the results of $L_{\phi}^{\rm TDUCF}$ inferred using the
saturated noise crossover function shows that incorrectly using the
unsaturated noise crossover function will result in inferred coherence
lengths less than the actual value.  A comparison of integrated noise
power amplitude to MFUCF resistance variances calculated from
magnetofingerprint data confirms that the Ag noise is not yet saturated.
The TDUCF noise in sample F would need to be integrated over 190 
decades in frequency (completely unphysical) to achieve the 
conductance fluctuation size seen in MFUCF.

\begin{figure}[h!]
\begin{center}
\includegraphics[clip, width=7.5cm]{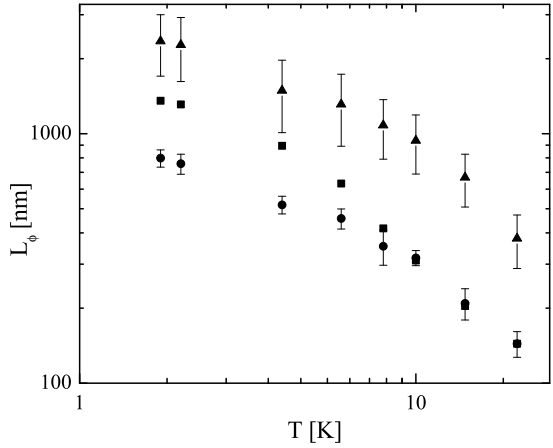}
\end{center}
\vspace{-3mm}
\caption{Coherence lengths from sample F.  Squares are from WL, circles are inferred via the unsaturated TDUCF crossover function, and triangles are calculated with the saturated TDUCF crossover function.}
\label{revfig5}
\end{figure}

Fig.~\ref{revfig5} shows the results of trying to infer $L_{\phi}^{\rm
TDUCF}(T)$ using the crossover function appropriate for saturated
TDUCF, in comparison with the weak localization data and the
$L_{\phi}^{\rm TDUCF}$ values calculated using the unsaturated TDUCF
expression.  Clearly the system is not in the saturated regime over
the observed temperature range.  However, the saturated crossover
function data {\it becomes a better match} to the WL data as $T$ decreases.
It seems reasonable that an interpolating crossover function between
the saturated and unsaturated crossover functions could be necessary.
If an unsaturated/saturated transition is at work, the coherence
lengths extracted via a correctly derived interpolating crossover
function could agree with those inferred from WL for all temperatures.
A definitive test of this hypothesis is to measure TDUCF and MFUCF in
extremely thin Ag samples down to dilution refrigerator temperatures
and compare their magnitudes and correlation fields.  These attempts
are ongoing.

\subsection*{C.  Drive dependence of the TDUCF}

We have also considered whether the nonequilibrium nature of the
transport measurements could result in the discrepancy seen in
Fig.~\ref{revfig4}.  As has been discussed extensively in
Ref.~\cite{Ovadyahu01PRB}, once a system is driven out of equilibrium,
it is a subtle question whether one should expect consistency between,
{\it e.g.}, $L_{\phi}^{\rm WL}(T)$, $L_{\phi}^{\rm TDUCF}(T)$, and
$R(T)$.  We have examined the drive dependence of our WL and 
TDUCF measurements, and return to this issue below.

\begin{figure}[h!]
\begin{center}
\includegraphics[clip, width=7.5cm]{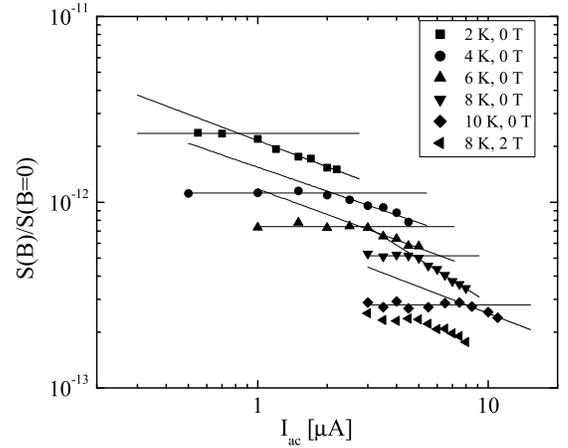}
\end{center}
\vspace{-3mm}
\caption{Sensitivity of the noise power of sample E to applied
current.  Enhanced suppression of the noise occurs when $eV_{\rm c}$
becomes larger than the correlation energy.  The drop-off current at
8~K, 2~T is the same as 8~K, 0~T implying that the coherence length is
not sensitive to large applied fields.}
\label{revfig6}
\end{figure}

Increasing the measurement current has a significant effect on the
measured noise power amplitude long before any change is observed in
$R(T)$, $L_{\phi}^{\rm WL}(T)$, or $L_{\phi}^{\rm TDUCF}(T)$.
This was seen previously in quasi-2D Ag films\cite{HoadleyetAl99PRB}.
A plot of noise power
versus applied current is given in Fig.~\ref{revfig6} for several
different temperatures.  It is clear that the noise power begins to
drop at a different applied current for each temperature.  

\begin{figure}[h!]
\begin{center}
\includegraphics[clip, width=7.5cm]{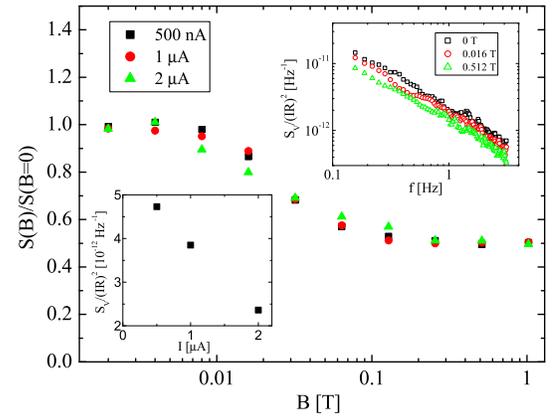}
\end{center}
\vspace{-3mm}
\caption{Normalized noise power of sample J at 2~K as a function of
field for three different values of drive current (rms 500~nA,
1~$\mu$A, and 2~$\mu$A).  The unchanging crossover field demonstrates
explicitly that the inferred $L_{\phi}^{\rm TDUCF}$ is {\it not}
affected strongly by drive level.  Left inset: dimensionless
conductance noise power as a function of drive current at zero field,
showing that the higher drive currents do suppress the {\it magnitude}
of the TDUCF.  Right inset: log-log plot of the normalized noise power
as a function of frequency at 2~K, 2~$\mu$A drive, for three different
values of $B$, showing the $1/f$ dependence typical for all the TDUCF
data in this work.}
\label{revfig7}
\end{figure}

We reemphasize that while the TDUCF amplitude is strongly affected by
drive level, the {\it field-dependence} used to infer $L_{\phi}^{\rm
TDUCF}$ is essentially unchanged (within the error bars) below drive
levels where heating is clearly manifest.  Figure \ref{revfig7} shows
the experimentally observed noise crossover data for sample J at 2~K
for three different drive currents.  The left inset indicates that the
noise power amplitude shows a distinct drive current dependence while
the crossover field of the three curves remains {\it unchanged}
(within the error bars).  Over the same range of drive currents,
neither $L_{\phi}^{\rm TDUCF}$ or $L_{\phi}^{\rm WL}$ are altered, to
within the error bars on those quantities.  This result implies that
the coherence length differences observed in the Ag samples (as in
Fig.~\ref{revfig4}b) are {\it not} a result of the type of
out-of-equilibrium dephasing effects described in
Ref.~\cite{Ovadyahu01PRB}.

A reasonable explanation for the decrease in TDUCF amplitude with
increasing drive current is energetic ensemble averaging as $eV_{\rm
c}$ becomes larger than the correlation energy of the samples.  As $T$
is increased, the decrease in $L_{\phi}$ and corresponding increase in
$E_{\rm c}$ would require larger drives to observe such averaging,
consistent with what is observed.  The inability to observe such a
suppression of noise at high drives in the AuPd samples, where
$L_{\phi}$ is much shorter, further supports this explanation.  An
estimate of the current required such that $eV_{\rm c} \sim E_{\rm c}$
in AuPd leads to a current greater than that needed empiricially to
heat the samples significantly.  

The detailed above-threshold dependence of the noise power on the
applied current, however, is surprising.  Properly normalized noise
power is not proportional to $1/I_{\rm ac}$ above the critical current
where $e \times I (R/L)L_{\phi} \sim E_{\rm c}$, as a simple treatment
would predict.  Instead the noise power decreases like $I_{\rm
ac}^{-0.5}$ above the threshold.  For completeness, the drive
dependence at 8 K was repeated at 2 T.  The suppression of the noise
amplitude started at the same threshold drive current, further
evidence that the coherence length in Ag does not change at high
fields ({\it i.e.} dephasing in Ag is not magnetic impurity scattering
in origin).

\subsection*{D.  Dephasing mechanisms}

The $L_{\phi}^{\rm WL}$ and $L_{\phi}^{\rm TDUCF}$ data have
implications for the decoherence mechanisms at work in these
materials.  As we argue below, the data strongly support that magnetic
impurity scattering is relevant in AuPd samples, and that scattering
from dynamical defects such as tunneling two-level systems (TLS) are
unlikely to be significant in these materials.

We first consider AuPd, in which the $L_{\phi}$ values at low
temperatures are significantly lower than those predicted from Nyquist
scattering.  Exact quantitative agreement between the theoretical
Nyquist length and experimental coherence length is not necessarily
expected since AuPd is not a simple metal, i.e. it does not have the
typical spherical Fermi surface.  However, if the Nyquist dephasing
mechanism is at work, one should expect that the experimental
coherence lengths at low temperature would show the predicted power
law dependence with the temperature.  Deviations from this power law
are particularly clear as apparent low-$T$ saturation of $L_{\phi}$ in
samples C and D, shown in more detail in Ref.~\cite{TrionfietAl04PRB}.

\begin{figure}[h!]
\begin{center}
\includegraphics[clip, width=7.5cm]{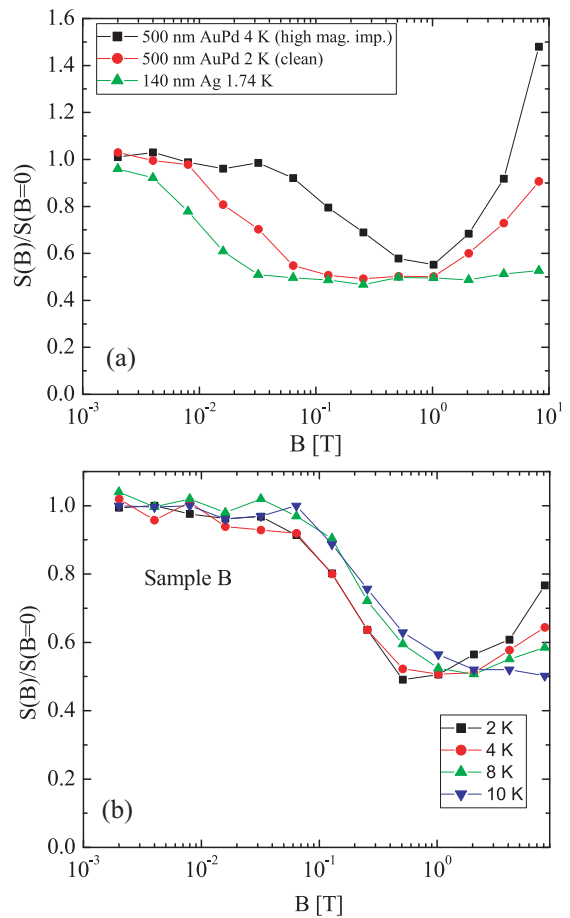}
\end{center}
\vspace{-3mm}
\caption{(a) High field noise data for samples C, D, and F, all at 2~K.  The upturn at high fields implies that non-negligible concentrations of magnetic impurities exist in the AuPd samples. (b) High field noise data for sample B at several temperatures, showing the upturn in this nominally clean AuPd sample.  The lines connecting the points for each temperature are a guide to the eye.  Note that the upturn is smaller and happens at higher fields for higher temperatures, consistent with Zeeman splitting of magnetic impurities.}
\label{revfig8}
\end{figure}

These deviations are due at least in part to detectable concentrations
of magnetic impurities in all AuPd samples.  The presence of such
impurities is strongly supported by high magnetic field noise power
data.  Figure~\ref{revfig8}a shows the normalized noise power versus
applied field up to 8 T for three samples, the AuPd sample
intentionally contaminated with magnetic impurities (D), one nominally
clean AuPd sample (C), and one Ag sample (F).  A noticeable upturn in
the noise power at high fields is seen in both AuPd samples, while
no such upturn is seen in the Ag at the lowest measured temperature.

The upturn is caused by a suppression of spin-flip decoherence as the
Zeeman splitting of the magnetic impurities exceeds $k_{\rm B}T$.  An
analogous upturn has been observed in investigations of Li
wires\cite{MoonetAl97PRB} and in recent Aharonov-Bohm measurements in
Cu rings\cite{PierreetAl02PRL}.  The increased coherence length leads
to an increase in the magnitude of the noise power via reduced
ensemble averaging.  Some upturn is visible at the highest $B/T$ ratio
in {\it all} AuPd samples, consistent with some magnetic impurities
even in nominally ``clean'' devices.  Figure~\ref{revfig8}b shows the
large field noise upturn at multiple temperatures in sample B.  The
characteristic field scale for the noise upturn increases with
increasing temperature, consistent with Zeeman splitting splitting of
magnetic scatterers.

 
One can use Eq.~(\ref{eq:nagaokasuhl}) to estimate the magnetic
impurity concentration required to produce the observed $L_{\phi}$
saturation values.  Assuming that the 2~K data represent complete
saturation (as appears to be the case for, {\it e.g.}, sample C), that
$T_{\rm K} \sim$~2~K, and a typical $\nu(E_{\rm F})$ for the noble
metals, one finds $c_{\rm mag} \sim$ 17~ppm for sample C.  This value
is surprisingly high and not consistent with the starting purity of
the source material.  One possible explanation for this would be an
enhancement of the spin-flip scattering process due to the strong
paramagnetism of the Pd component of the host alloy.  Note that even
if this concentration of magnetic impurities is accurate, the inferred
$T^{*}$ below which spin-flips should affect UCF and WL differently is
$<$~1~K, outside the range of these measurements.

Comparing the saturated values of $L_{\phi}$ for two samples allows
the {\it relative} concentrations of magnetic impurities to be
computed, independent of possible paramagnetic enhancement or density
of states uncertainties, since $\frac{c_{\rm mag,2}}{c_{\rm
mag,1}}=\frac{D_{2}}{D_{1}}\left(\frac{L_{\phi,1}}{L_{\phi,2}}\right)^{2}$.
Comparing sample C (quasi-2d) and D (quasi-2d, ``spiked'' with
additional magnetic impurities), one finds that $c_{\rm mag,D}/c_{\rm
mag, C} \approx 5$.  We note that $R(T)$ for these two samples shows
no discernable difference beyond what would be expected from their
diffusion constants.  This also supports the hypothesis that the
absolute concentration above is an {\it overestimate}, since
$\sim$~100~ppm magnetic impurities in sample D would likely cause
other discernable effects in addition to enhanced dephasing.
Furthermore, at such a concentration the crossover temperature $T^{*}
\sim$~4~K; however, no change in $L_{\phi}^{\rm WL}$
vs. $L_{\phi}^{\rm TDUCF}$ is apparent there, again suggesting that
this concentration is an overestimate.

While the above data show that spin-flip scattering is definitely
relevant in AuPd samples, unconventional (non-Nyquist, non-spin-flip)
dephasing must also be considered as an alternative explanation for
the suppression of low-$T$ coherence lengths as compared to those
expected from Nyquist theory.  Coherence lengths inferred via WL in Ag
samples (G \& H) in dilution refrigerator measurements show some
evidence of saturation at temperatures much less than 1~K.  (Other Ag
samples were not measured at dilution temperatures, and clearly had
not saturated down to 1.7~K.)  The cause of this saturation cannot be
dismissed easily as spin-flip scattering at this stage, since we have
yet to perform measurements ({\it e.g.}  TDUCF, Aharonov-Bohm) that
would directly probe for magnetic impurities in these samples at those
temperatures. 

Another proposed mechanism put forth to explain observed coherence
saturation is a dephasing mechanism caused by the same dynamic
two-level systems that cause
TDUCF\cite{ZawadowskietAl99PRL,ImryetAl99EPL}.  The data presented
herein imply that such a mechanism is unlikely to be the cause of the
coherence saturation seen in the AuPd.  Using the TDUCF noise
amplitude of the two materials at a given temperature, it is possible
to use the results of Feng, Lee, and Stone\cite{FengetAl86PRL} and the
measured sample parameters to estimate the ratio of TLS in the two
materials.  If the microscopic scattering properties of the TLS are
assumed to be identical in AuPd (sample A) and Ag (sample J), using
the appropriately normalized TDUCF magnitude at 2~K, we find that the
density of TLS is $\sim$ three times larger in Ag than in AuPd.  This
number should be considered a rough estimate since the scattering
cross-sections of the TLS in the two materials may be different.  A
conclusion that the mobile defect density is roughly the same in both
materials is reasonable.  If TLS-induced dephasing truly is
significant in AuPd at $\sim$ 2~K, one would therefore expect it to be
similarly important in Ag samples.  Given the excellent agreement at
that temperature of $L_{\phi}^{\rm WL}$ in Ag with the Nyquist
prediction, this seems unlikely.

\begin{table}
\caption{$L_{\phi}^{\rm WL}$(max) normalized by
that for sample H, compared with the expected ratio
from Eq.~(\ref{eq:GZ}) for zero-temperature saturation
of $\tau_{\phi}$, using sample parameters from Table~\ref{tab1}.}
\begin{tabular}{c c c c}
\hline
\hline
Sample ratio & Measured & Uncertainty & Expected \\
\hline
C/H & 0.0597 & 0.0017 & 0.0148 \\
D/H & 0.0221 & 0.0006 & 0.0218 \\
G/H & 0.6561 & 0.0186 & 0.7502 \\
\hline
\hline
\end{tabular}
\label{tab2}
\end{table}

For completeness, we compare maximum (lowest $T$) coherence
times observed for both AuPd and Ag with a prediction of
zero temperature coherence time saturation\cite{GolubevetAl02JLTP}.
The predicted zero-temperature dephasing time is
\begin{equation}
\frac{1}{\tau_{\phi,{\rm sat}}} = \frac{\sqrt{2}\rho e^{2}}{3 h \pi \sqrt{D}}\left(\frac{b}{\tau_e}\right)^{3/2}.
\label{eq:GZ}
\end{equation}
Here, $\tau_{e}$ is the elastic relaxation time of the conduction
electrons, $\rho$ is the resistivity, and $b$ is a constant of order
one.  Since a strict value for $b$ is not known, we select one sample
as a reference, and compare the ratio of each maximum
$L_{\phi}^{\rm WL}$ to the ratio expected from Eq.~(\ref{eq:GZ}) and
the measured resistivities.  We acquired WL data on Ag samples G and H
in the dilution refrigerator, with some indications of $L_{\phi}^{\rm
WL}$ saturation; for AuPd samples C and D, such saturation has an
onset at temperatures above 2~K.  Since the Ag shows the least
magnetic contamination, we choose sample H as the reference.  In
Table~\ref{tab2}, we show the predicted ratios from applying
Eq.~(\ref{eq:GZ}), which fall well outside the errors in the
experimentally measured ratios for two of the three samples.  In fact,
the one sample that shows good agreement with (\ref{eq:GZ}) is sample
D, the sample intentionally contaminated with {\it additional}
magnetic impurities known to significantly impact coherence.  These
data do not appear to support Eq.~(\ref{eq:GZ}).  Given the
demonstrated effectiveness\cite{PierreetAl03PRB} of trace magnetic
impurities to affect $L_{\phi}^{\rm WL}$, one must view analyses of
low temperature coherence saturation with appropriate caution.


\section*{V.  Conclusion}

We have performed detailed measurements of the WL magnetoresistance,
TDUCF amplitude as a function of field, and the magnetofingerprint of
both AuPd and Ag mesoscopic wires.  A comparison of the coherence
lengths inferred from WL and TDUCF measurements are in strong
quantitative agreement for the AuPd samples while a disagreement was
observed in the Ag samples similar to that previously seen in Ag
films.  We hypothesize that the reason for this discrepancy is that Ag
may be approaching the saturated noise limit, and further experiments
to test this are ongoing.  The observed suppression of TDUCF magnitude
at high drive currents in Ag agrees qualitatively with ensemble
averaging related to the correlation energy scale, though a
quantitative understanding is still lacking.  Finally, we have
considered the coherence saturation seen in our quasi-2D AuPd samples,
and discussed the important influence of magnetic impurities in such
systems.  The subtle physics, rich phenomenology, and continued
presence of surprises (such as the disagreement between WL and TDUCF
in Ag) demonstrate why electronic coherence in solids remains a lively
area for investigation.

We would like to thank N.O. Birge for his helpful advice concerning
noise measurements, and I.L. Aleiner and A.D. Stone for discussions of
the theory.  This work was supported by DOE grant
DE-FG03-01ER45946/A001 and the David and Lucille Packard Foundation.





\end{document}